\documentclass[aps,a4paper,preprint,superscriptaddress,nofootinbib]{revtex4-1}

\usepackage{pdfpages}
\usepackage{graphicx}

\begin{document}

\title{Worldwide spreading of economic crisis}

\author{Antonios Garas}
\email{agara@physics.auth.gr}
\author{Panos Argyrakis}
\affiliation{Department of Physics, University of Thessaloniki, 54124 Thessaloniki, Greece.}
\author{C\'{e}line Rozenblat}
\affiliation{Geography Institute, Faculty of Geosciences, University of Lausanne, 1015 Lausanne, Switzerland.} 
\author{Marco Tomassini}
\affiliation{Faculty of Business and Economics, University of Lausanne,\\ 1015 Lausanne, Switzerland.}
\author{Shlomo Havlin}
\affiliation{Minerva Center and Department of Physics, Bar-Ilan University,\\ 52900 Ramat Gan, Israel.}

\begin{abstract}
\bf We model the spreading of a crisis by constructing a global economic network and applying the Susceptible-Infected-Recovered (SIR) epidemic model with a variable probability of infection. The probability of infection depends on the strength of economic relations between the pair of countries, and the strength of the target country. It is expected that a crisis which originates in a large country, such as the USA, has the potential to spread globally, like the recent crisis. Surprisingly we show that also countries with much lower GDP, such as Belgium, are able to initiate a global crisis. Using the {\it k}-shell decomposition method to quantify the spreading power (of a node), we obtain a measure of ``centrality'' as a spreader of each country in the economic network. We thus rank the different countries according to the shell they belong to, and find the 12 most central countries. These countries are the most likely to spread a crisis globally. Of these 12 only six are large economies, while the other six are medium/small ones, a result that could not have been otherwise anticipated. Furthermore, we use our model to predict the crisis spreading potential of countries belonging to different shells according to the crisis magnitude.

\end{abstract}


\maketitle

\section{Introduction}

A global economic crisis, such as the recent 2008-2009 crisis, is certainly due to a large number of factors. In today's global economy, with strong economic relations between countries, it is important to investigate how a crisis propagates from the country of origin to other countries in the world. Indeed, several significant crises in the past few decades have been originated in a single country. However, it is still not clear how and to what extent domestic economies of other countries may be affected by this spreading, due to the inter-dependence of economies ~\cite{Krugman}. Here, we use a statistical physics approach to deal with the modern economy, as it has been done successfully in the recent years for the case of financial markets and currencies~\cite{MantegnaStanley,GGPS,Mantegna1999,Onnella2004,Garas1,Garas2,Garas3,Richmond2007,Takayasu2007}. More precisely, we view the global economy by means of a complex network~\cite{BarabasiAlbert,Mendes,Hidalgo}, where the nodes of the network correspond to the countries and the links to their economic relations.
 
For generating the economic network we use two databases, in order to avoid any bias due to the network selection. A global Corporate Ownership Network (CON) is extracted from a database of the 4000 world corporations with the highest turnover, obtained from the {\it Bureau Van Dijk}~\footnote{Bureau van Dijk Electronic Publishing (BvDEP) http://www.bvdep.com/}. This database includes all the corporate ownership relations to their 616000 direct or indirect subsidiaries for the year 2007. 
The trade conducted by these companies, in terms of import/export, is a large fraction of the total world trade. Furthermore, the network of subsidiaries is a direct measure of the investment of large corporations in order to grow. Foreign investment is a key factor for the development of global and local economies while, as recent economic crises suggest, the role of large corporations to the spreading of crisis in the global economy is yet not clearly understood. 
The second network, the International Trade Network (ITN), is extracted by the 2007 version of the CHELEM database obtained by {\it Bureau Van Dijk}~$^1$, which contains detailed information about international trade, and GDP values for 82 countries in million US dollars. This database provides us with an economic network based on import/export relations between countries. 

For both networks we are able to locate a nucleus of countries that are the most likely to start a global crisis, and to sort the remaining countries crisis spreading potential according to their ``centrality''. Initially, a crisis is triggered in a country and propagates from this country to others. The propagation probability depends on the strength of the economic ties between the countries involved and on the strength of the economy of the target country. Our results show that, besides the large economies, even smaller countries have the potential to start a significant crisis outbreak.

The CON is a network that connects 206 countries around the globe, using as links the ownership relations within large companies. If companies listed in country A have subsidiary corporations in country B, there is a link connecting these two countries directed from country A to country B. The weight of the link, $w_{AB}$, equals the number of the subsidiary corporations in country B controlled by companies of country A. Next, if companies from country B have subsidiary corporations in country C, then again there is a weighted link, $w_{BC}$, connecting these two countries directed from B to C, and so on. This way we obtain a network with total 2886 links among 206 nodes (countries). Of these links 685 are bi-directional, meaning that if there is a link from node $i$ to $j$, as well as a link from node $j$ to $i$, and the rest 1516 are one directional only. 

We assume that the total link weight between a pair of nodes (countries) $ij$ is the sum of all links independently of their direction, $w^{(ij)}_{\rm tot}=w_{ij}+w_{ji}$. The total link weight represents the strength of the economic ties between two countries in the network. We quantify the total economic strength of a country $i$ by its total node weight, $\tilde{w}^{i}_{\rm tot}=\sum_{j}{w_{ij}}+\sum_{j}{w_{ji}}$, i.e., summing the weights of all links of node $i$. The probability density distributions of the total node weights and of the total link weights is skewed and heavy tailed, as shown in Fig.~S1 in the Supplementary Information. We find an almost linear relation between $\tilde{w}^{i}_{\rm tot}$ and the GDP of country $i$, (as shown in supplementary Fig.~S2) which indicates that the total weight of a country in our network is strongly correlated to a traditional economic measure.

The ITN is calculated from the second database after we aggregate the trade actions between all pairs of countries. Using the trading relations between each pair of countries e.g., A and B, we can create a bi-directional network where $E_{AB}$ represents the export of A to B, and $E_{BA}$ represents the export of B to A. Of course $E_{AB}$ is equal to $I_{BA}$, which stands for the imports of B from A. In accordance to the above notations, the total link weight is given by $w^{(ij)}_{\rm tot}=E_{ij}+E_{ji}$, but the total node weight $\tilde{w}^{i}_{\rm tot}$ which quantifies the economic strength of a node equals to its GDP value. 

\begin{figure}[!ht]
\begin{center}
\includegraphics[width=8cm]{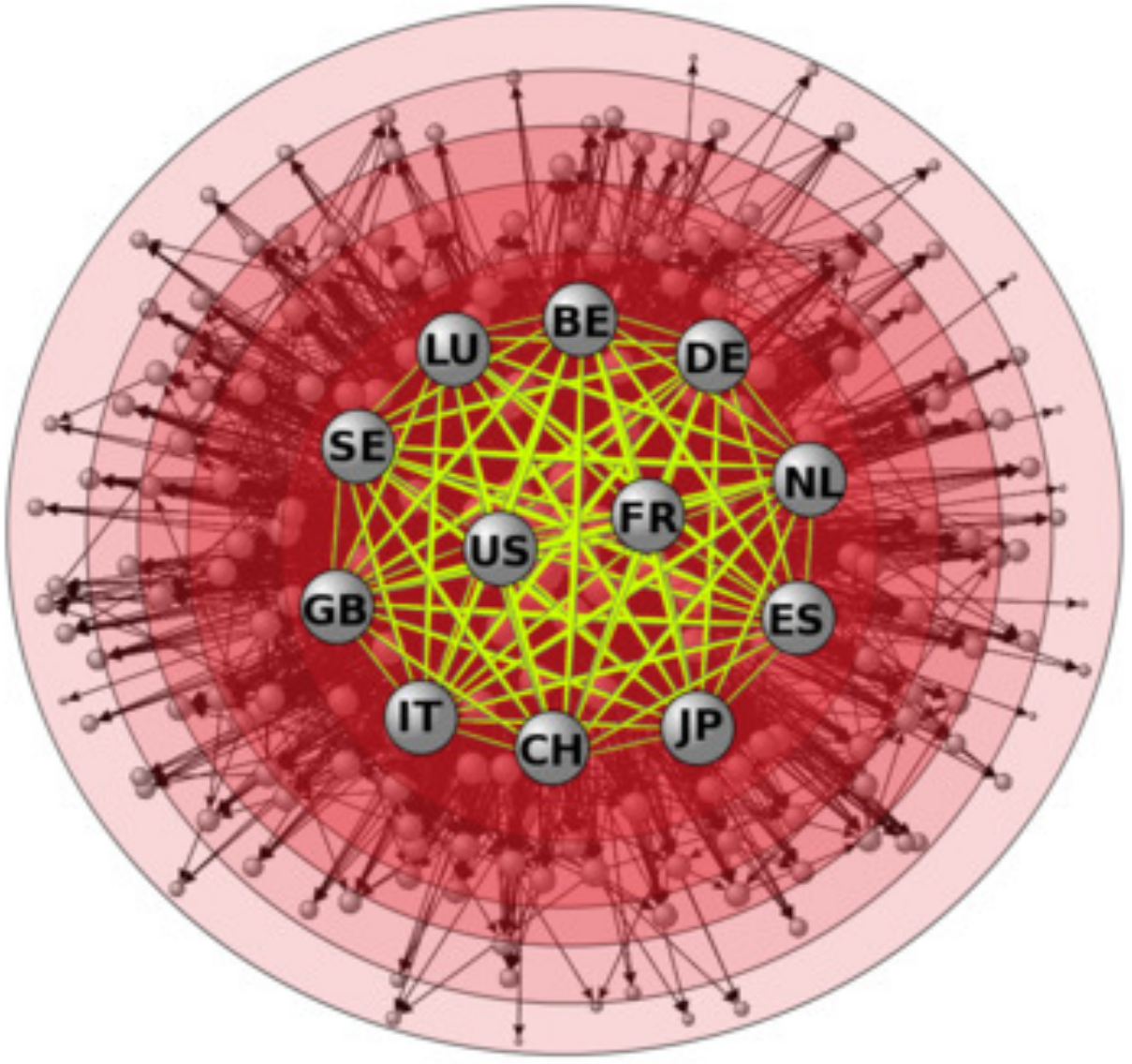}
\end{center}
\caption{
{\bf  An illustration of the layered structure of the global economic network of 206 countries of the world using the large corporation subsidiary relations.} The layers are a schematic drawing of shells obtained by the $k$-shell method. The outer layers include the loosely connected countries, while at the center we highlight the nucleus of the 12 countries we identified.
}
\label{fig1}
\end{figure}

To identify the uneven roles of different countries in the global economic network, we use the $k$-shell decomposition and assign a shell index, $k_{s}$, to each node. The $k$-shell is a method identifying how central is a node in the network, the higher its $k_{s}$ the more central role the node is considered to have as a spreader~\cite{Pittel,Carmi,Seidman}. The nodes in the highest shell, called the nucleus of the network, represent the most central countries. To determine the $k$-shell structure we start by removing all nodes having degree $k=1$, and we repeat this procedure until we are left only with nodes having $k\geq 2$. These nodes constitute shell $k_{s}=1$. In a similar way, we remove all nodes with $k=2$ until we are left only with nodes having degree $k\geq 3$. These nodes constitute $k_{s}=2$. We apply this procedure until all nodes of the network have been assigned to one of the $k$-shells. With this approach we view the network as a layered structure. The outer layers (small $k_{s}$) include the loosely connected nodes at the boundary of the network, while in the deeper layers we are able to locate nodes that are more central. An illustration of this structure is shown in Fig.~\ref{fig1}.

\begin{figure}[!ht]
\begin{center}
\includegraphics[width=15cm]{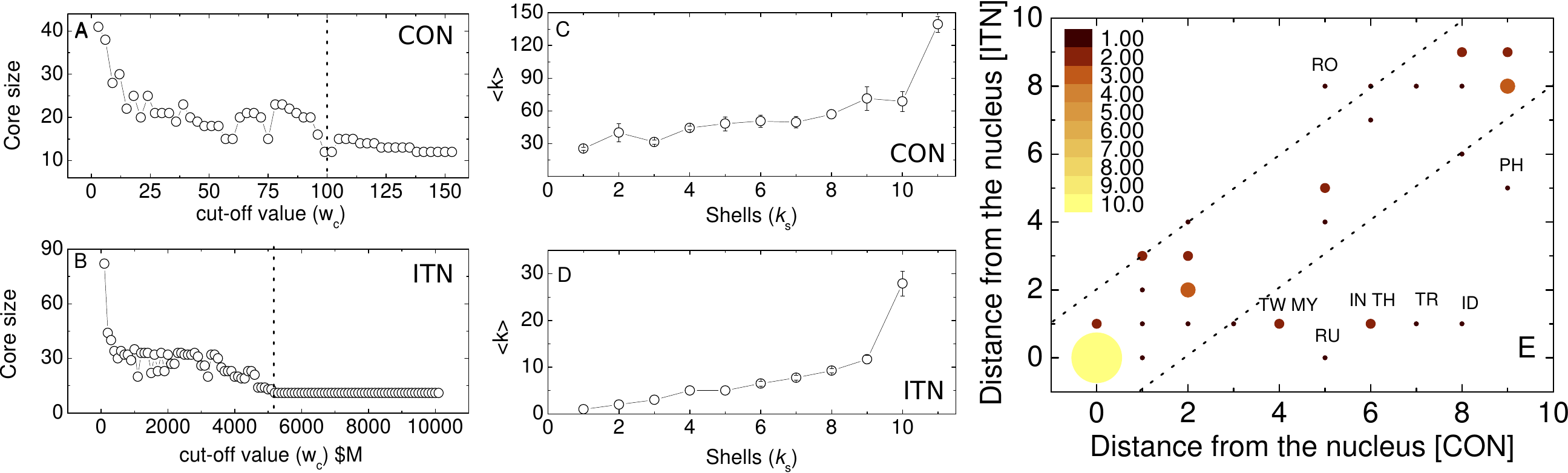}
\end{center}
\caption{
{\bf  $k$-shell decomposition of the network.} (A) The size of the nucleus for different cut-off values, using CON. (B) The size of the nucleus for different cut-off values, using the ITN. (C) The average degree of the countries of shell $k_{s}$ for CON.The higher the shell number, the closer we get to the nucleus of the network. This plot corresponds to the structure obtained for the cut-off value $w_{c}=100$. (D) The average degree of the countries of shell $k_{s}$ for the ITN. This plot corresponds to the structure obtained for the cut-off value $w_{c}=5100 \$$M. (E) Scatter plot showing the distance from the nucleus of different countries for both CON and ITN.}

\label{fig2}
\end{figure}

To identify the nucleus of CON we consider in the $k$-shell the network having only links with weights above a cut-off threshold $w_{c}$. By using different threshold values we locate different nuclei of different sizes, as shown in Fig.~\ref{fig2}(A). However, for the whole range of the threshold values used, namely for $w_{c}\in[0,150]$, we are able to identify twelve countries that are {\it always} present in the nucleus with $k_{s}=11$. Furthermore, for $w_{c}\geq 100$ the nucleus is fully connected and always include only the same twelve countries. These countries are the USA (US), United Kingdom (GB), France (FR), Germany (DE), Netherlands (NL), Japan (JP), Sweden (SE), Italy (IT), Switzerland (CH), Spain (ES), Belgium (BE), and Luxembourg (LU), sorted according to their total node weight $\tilde{w}_{\rm tot}$.

When performing the $k$-shell decomposition method on the ITN we locate a nucleus at $k_{s}=10$ composed of 11 countries, that is stable and always the same for  $w_{c} \geq 5100 \$$M as shown in Fig.~\ref{fig2}(B). Actually we may view it as a 12 country core, because the CHELEM database regards Belgium and Luxembourg as one trade zone, and therefore the core consists of the following twelve countries: China (CN), Russia (RU), Japan (JP), Spain (ES), United Kingdom (GB), Netherlands (NL), Italy (IT), Germany (DE), Belgium (BE), Luxembourg (LU), USA (US), and France (FR). This  nucleus is almost the same as the nucleus found for CON. There are only two differences, which can be clearly understood  due to the complimentary nature of the networks. The first difference is that Sweden (SE) and Switzerland (CH) are now located at shell 9 in ITN, which is one shell below of the core. The second difference is that China (CN) and Russia (RU) are now part of the nucleus, while in the CON they were located one shell and five shells away from the core, respectively. The presence of these countries in the nucleus of ITN, and their absence in the nucleus of CON can be explained considering the unusual structure of their economies. There are not many large corporations in the global corporate database having Headquarters in China or in Russia, because most of the goods exported by these countries (especially by China) are branded under Western brand names. Therefore most of their trade comes from subsidiaries of Western corporations. On the other hand their total trade volume is significantly high in the total global trade.

Compared to the other shells, the countries in the nucleus are not only strongly interconnected among themselves, but also have many links to other nodes of the network. This is clearly demonstrated in Fig.~\ref{fig2}(C) for CON and in Fig.~\ref{fig2}(D) for the ITN. More specifically, for CON the 12 countries of the nucleus have significantly large average degree $\left< k \right>=139 \pm 7$ compared to the countries in the shells below, like  $\left< k \right>=69 \pm 9$ for shell $k_{s}=10$. For ITN the average degree of the nucleus is $\left< k \right>=28 \pm 2$, while the average degree of shell $k_{s}=9$ is $\left< k \right>=12 \pm 1$. 
Surprisingly, not all 12 countries have the largest total weights or the largest GDP. Nevertheless, our results suggest that they do play an important role in the global economic network. For CON i.e. we find that six of the G8 members are part of the nucleus (except Russia and Canada that belong to lower shells), while the other six are smaller countries in absolute size (relative to the large countries). This is explained by the fact that these smaller countries do not support only their local economy, but they are a haven for foreign investments, as they attract funds from large countries for taxation purposes, safekeeping, etc. and a problem in such investments can easily lead to a chain reaction in other countries. Countries such as LU and CH, which are headquarters for some of the world's largest companies and subsidiaries, interact very strongly with a very large number of countries. For example, about 95\% of all pharmaceutical products of the Swiss industry are not intended for local consumption, but for exporting.

Although, both CON and ITN are based on different databases, it is interesting that most of the countries are located in very close $k$-shells, supporting the robustness of our approach. This is shown in the scatter plot of Fig.~\ref{fig2}(E) which maps the change in the $k$-shell ranking for both networks. The ranking is done in units of distance from the nucleus. We find that most of the countries are located in almost the same distance from the nucleus for both networks, since 82\% of the countries are inside the two lines (Fig. 2(E)), representing distance $\leq 1$. The few countries that are in very different $k$-shells are usually large countries (in terms of population), which are very active in terms of trade but do not own many large corporations with global status, i.e, Turkey (TR) and India (IN).

It should be noted that when we apply the clique percolation method (CPM)~\cite{Palla}, both CON for $w_{c}\geq 100$ and ITN for $w_{c} \geq 5100 \$$M, we obtain that the strongest connected communities are the same as the nuclei found using the $k$-shell method, as shown in supplementary Fig.~S3.

Next we study how an economic crisis can propagate on these networks. An economic crisis is a very complex phenomenon that cannot be reduced to an ``all or nothing'' situation. In order to get some insight on the mechanisms of crisis spreading and on the role of the network topology in this spreading, we applied a Susceptible-Infected-Recovered (SIR)-type model. The SIR model has been used successfully to model spreading of epidemics in various networks~\cite{Newman,Colizza}. The basic characteristic of SIR is that it usually assumes a fixed probability for neighbour-to-neighbour infection. In our case we assume a probability which depends on the economic weights of the links and the strength of the targeted country (see Eq.~\ref{eq1}). Initially all nodes are in the susceptible (S) state. We chose one node (country) and set it to be in a crisis (infected) state (I). During the first time step it will infect all its neighbouring nodes with probability calculated from Eq.~\ref{eq1}, and the status of the infected nodes switches from S to I. This process is repeated with all infected nodes trying to infect during each time step their susceptible (S) neighbours. After each time step, the status of the original infected nodes changes to recovered (R) and can no longer infect or become infected. In our economic crisis epidemics the recovery means that a set of successful measures has been applied and the country overcomes the crisis. The simulation stops when there are no more infected nodes, or when all the nodes have been infected.

We assume for both, CON and ITN, that the probability $p_{ij}$ that node $i$ infects its neighbouring node $j$ depends on the total weight of the link, the total weight (strength) of the targeted node, and on the magnitude $m$ of the crisis, as follows
\begin{equation}
p_{ij}\propto m\cdot w^{(ij)}_{\rm tot}/\tilde{w}^{j}_{\rm tot}.
\label{eq1}
\end{equation}
The ratio $w^{(ij)}_{\rm tot}/\tilde{w}^{j}_{\rm tot}$ represents the relative economic dependence of country $j$ on country $i$, which we consider as a factor in the probability that country $j$ will be infected by country $i$. The factor $m$ represents the strength of the crisis and can obtain any positive value.

In economic relations directionality plays an important role. For CON i.e., a crisis in a subsidiary could create some problems in the mother company, but if there is a crisis in the mother company its effect on the subsidiaries is always significantly more severe. To account for directionality, we study the case when we keep only one link, directed or undirected, for each pair of connected nodes. To this end we calculate for all pairs of nodes the quantity
\begin{equation}
T=\left| \frac{w_{ij}-w_{ji}}{w_{ij}+w_{ji}} \right|.
\label{eq2}
\end{equation}
If $T$ is smaller than a certain threshold value, which is a parameter, then the link will be regarded as undirected. Otherwise, the link is directed pointing from $i$ to $j$ if $w_{ij}>w_{ji}$, and from $j$ to $i$ if $w_{ij}<w_{ji}$. In all cases the weight of the link is $w^{(ij)}_{\rm tot}=w_{ij}+w_{ji}$. In the current study we set the threshold value of $T$ to be zero. By increasing this value, we increase the percentage of undirected links in the resulting network, and if we set $T=1$ then the entire network is undirected. Our results are not sensitive to the value of T. We find that for all values of $T<0.5$ the crisis epidemics is nearly the same as shown in supplementary Fig.~S4. 

In Fig.~\ref{fig3} we present the results of the SIR simulations. Figs.~\ref{fig3}(A) and \ref{fig3}(B) show that a crisis propagates to larger parts of the world when it has larger magnitude. Note also that the directed network significantly delays the actual propagation of the crisis, in comparison to the random case (see legend) and to the undirected case. This means that if the directionality were not present, the crisis spreading would have been more severe. Figs.~\ref{fig3}(C) and \ref{fig3}(D) show how a crisis spreading process in CON depends on different origin countries according to their shell. We find that the countries in the nucleus can spread a crisis to larger parts of the world compared to countries in the outer shells, even if the crisis originates in a small country, such as BE. Applying the SIR model to the ITN yields the results of Figs.~\ref{fig3}(E) and \ref{fig3}(F). It is clear that from both figures we draw the same conclusion, i.e. the inner the shell, the more severe a crisis outbreak originating to this shell will be. Additionally, we show that in the ITN a crisis could have more severe effects, since the outer shells are capable of larger global impact in comparison to CON. It is interesting to note that the error bars in Figs.~\ref{fig3}(C) - \ref{fig3}(F) are small, which means that all the countries in the shell have similar behavior, and the shell determines the spreading instead of the country size. This indicates the predictability power of our method. 

In Fig.~\ref{fig4}(A) we show the world countries colored according to the shell they belong in CON, and in Fig.~\ref{fig4}(B) we show the countries affected by the economic crisis that started in 2008 in USA, which has spread globally. In Fig.~\ref{fig4}(C) we show a prediction of the model for CON. We simulate an infection of a crisis starting in the US, using $m=4.5$, which leads to roughly the same percentage of infected countries as the actual crisis of 2008-2009 ($\sim 90\%$ of the world countries reported economic slowdown due to the crisis). The percentage of correctly predicted countries infected by the crisis using the model is above the average percentage of the prediction using random selection. A quantitative agreement between the prediction power of our model and the actual infection strength is shown in Fig.~\ref{fig4}(D).

Considering the example of BE (ranked 29th according to its total GDP), we find that a crisis originating in this country with magnitude $m=4.5$ is able to affect for CON almost 60\% of the world countries (average result of 50 realizations), while the worst case scenario that is given by the maximum value of the fraction of infected countries (out of the same 50 realizations) is $95\%$ of global infection. For a crisis starting in the nucleus of both networks we find that this maximum fraction has a sharp transition (see suplementary Figs.~S5 and~S8), verifying that even crises with small magnitude can propagate to the rest of the world, if they originate in the inner shell. This shows the crucial role that these 12 countries play in the world economy. But surprisingly even countries of intermediate shells may play a major role in crisis spreading, as well. Such countries, as we show in Figs.~\ref{fig3}(C) and \ref{fig3}(D), are able to spread a crisis with sufficient magnitude to large fractions of the world as well. This is similar to what happened in the case of the Asian currency crisis, starting in Thailand in 1997. Thailand belongs to the $k_{s}=5$ shell for CON and to the $k_{s}=9$ shell for ITN, but a crisis originating there was able to affect countries of higher shells, and eventually triggered a new crisis that started in 1998 in Russia~\cite{Desai}.
 
\section{Discussion}

We model the way an economic crisis could spread globally, depending to the country of origin. We first create two global networks, which describe strong interaction patterns of the world economy. The first network (CON)  is based on the world's largest companies and all their subsidiaries and links together 206 countries, and the second network (ITN) is created using aggregated trade data linking together 82 countries. A crisis is triggered with a controlled magnitude and propagates from one country to another with a probability that depends on the strength of the economic ties between the countries involved and on the strength of the economy of the target country. Furthermore, using the $k$-shell decomposition method we are able to identify the role of the different countries in a world crisis, according to the shell they belong, and we identify the 12 most effective countries for crisis spreading. 

We find that although a recent global economic crisis originated in the USA, this might not be always the case and even smaller countries have the potential to start similar crisis outbreaks. In retrospect, we know that this has happened several times in the past (e.g. the crisis in Indonesia), and is happening these days as well, with a crisis in Greece~\footnote{Greece belongs to shell 5 in CON, and in shell 2 in the ITN.} threatening to cause an avalanche effect to other European economies. In part III of the supplementary information we examine in more detail the case study of Greece, where we show that the fears of contagion of other major European economies by the Greek crisis are justified. We report almost 40\% probability of infection for the major EU countries only through the contagion channels modeled using CON and ITN i.e. not taking into acount the loans Greece received from other countries. This is in contrast to the common belief that smaller countries partake only in crises that are locally contained, but do not spread out to affect the larger countries. Thus, our findings show the predicting power of the network approach using the $k$-shell ranking methodology. This analysis is important when establishing international commerce rules, policies and legislations, trade treaties and alliances, as they can have a serious impact on monetary policies and international affairs.


\begin{figure}[!ht]
\begin{center}
\includegraphics[width=15cm]{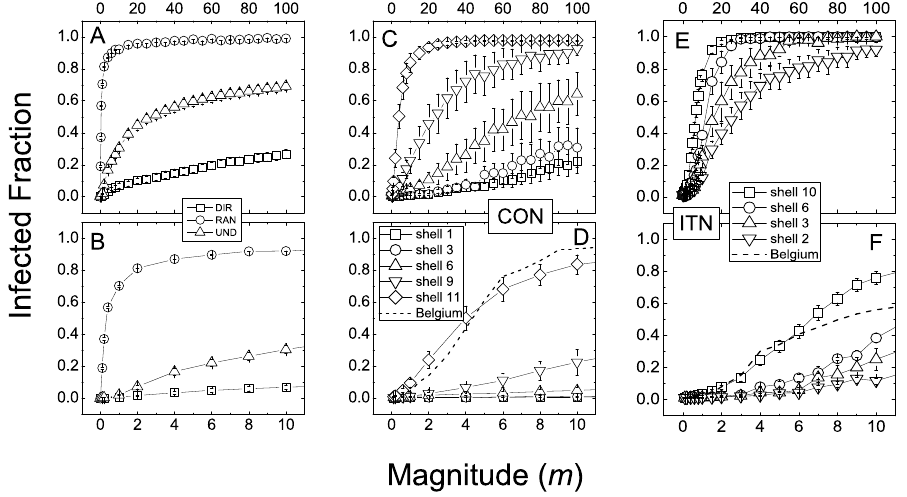}
\end{center}
\caption{
{\bf  Modeling the economic crisis propagation using SIR dynamics.} (A) Infected fraction of nodes infected by a crisis spreading versus the magnitude $m$ of the crisis. DIR is the real topology of CON, taking into account the directionality, UND is a network similar to the real one but undirected, and RAN is a simulated network with the same number of nodes, the same number of links per node, but with shuffled weights and directions. (B) Zoom of the area showing the spreading for smaller crisis magnitudes $m$. (C) Fraction of nodes infected by a crisis originating from different shells of the network versus its magnitude $m$ for CON. (D) Zoom of the area showing the spreading for smaller crisis magnitudes $m$. The dashed line shows the spreading of a crisis originated in Belgium, which is one of the smaller countries that belong to the nucleus of the network. Note that a crisis originating in BE, as $m$ gets larger, becomes more severe in comparison to the average case for all the countries in shell 11. (E) Fraction of nodes infected by a crisis originating from different shells of the network versus its magnitude $m$ for the ITN. (F) Zoom of the area showing the spreading for smaller crisis magnitudes $m$. The dashed line again shows the spreading of a crisis originated in Belgium. 
The results are averages over 50 realizations for each node of the network, and the error bars are showing the standard deviation.}
\label{fig3}
\end{figure}

\begin{figure}[!ht]
\begin{center}
\includegraphics[width=17cm]{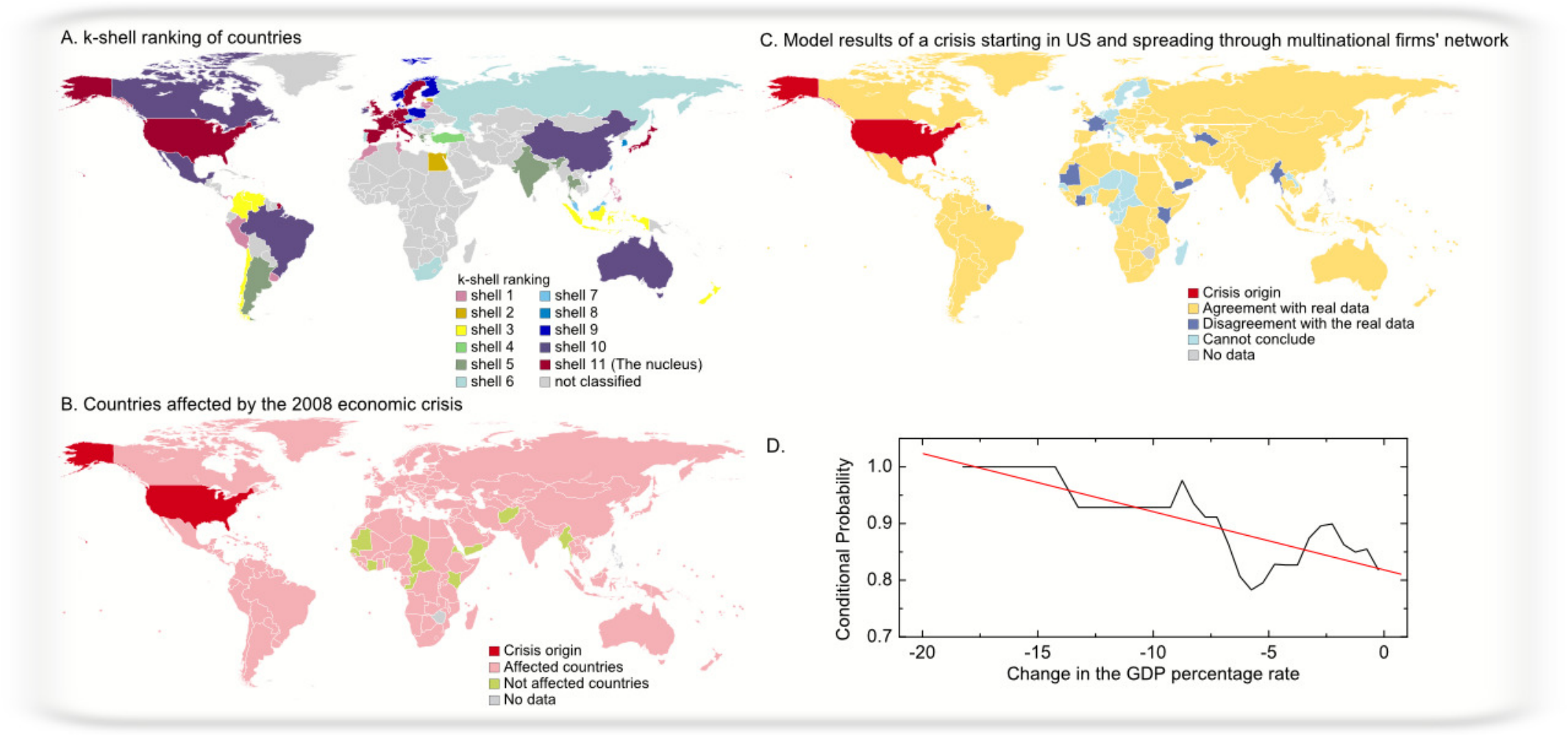}
\end{center}
\caption{
{\bf  Spreading of a real crisis to the world.} (A) The world countries coloured according to the $k$-shell they belong for CON. (B) The countries affected by the economic crisis which started in the US in 2008. The effect of the crisis is an economic slowdown, as it is reflected by the annual change of the GDP published by the International Monetary Fund. (C) Model results for CON, showing the crisis starting in the US and spreading to the rest of the world. We performed 1000 realizations always starting an infection in the US. For each realization, when the simulation ends, every country has a score = 1 if it is infected and 0 if it is not infected. A sum of the scores per country for all realizations leads to a number inside the interval $[0, 1000]$, where 0 means that in all runs this country was not infected and 1000 means that in all runs it was infected. We set a threshold value at 80\%, so if a country has score $\geq 800$ then it is considered as infected. If it has score $\leq 200$ then it is considered as not infected, and if it is in the range $(200, 800)$ we say that we cannot conclude about its status.  We find that the average percentage of infected countries is 90.6\%, while the worst case scenario which is given by the maximum number of infected countries in all our runs is 96.6\%. For comparison purposes if we start a crisis of the same magnitude in countries of lower shells we find a much lower percentage. For example, if we start the crisis in Russia (ks = 6) the average fraction of infected countries is 3.34\%, while the worst case scenario which is given by the maximum value of the fraction of infected countries is 18.9\%. (D) Probability of the correct prediction of the model as a function of the 2009 vs. 2008 change in the GDP percentage rate. The curve is smoothed using 6 point moving average. The trend that is shown by the linear fit (red curve) shows quantitatively that the more affected a country is by the real crisis of 2008 (represented by larger change in GDP), the higher the probability of our model to yeld a correct prediction.}
\label{fig4}
\end{figure}

\clearpage
\mbox{}



\section*{Supplementary material}

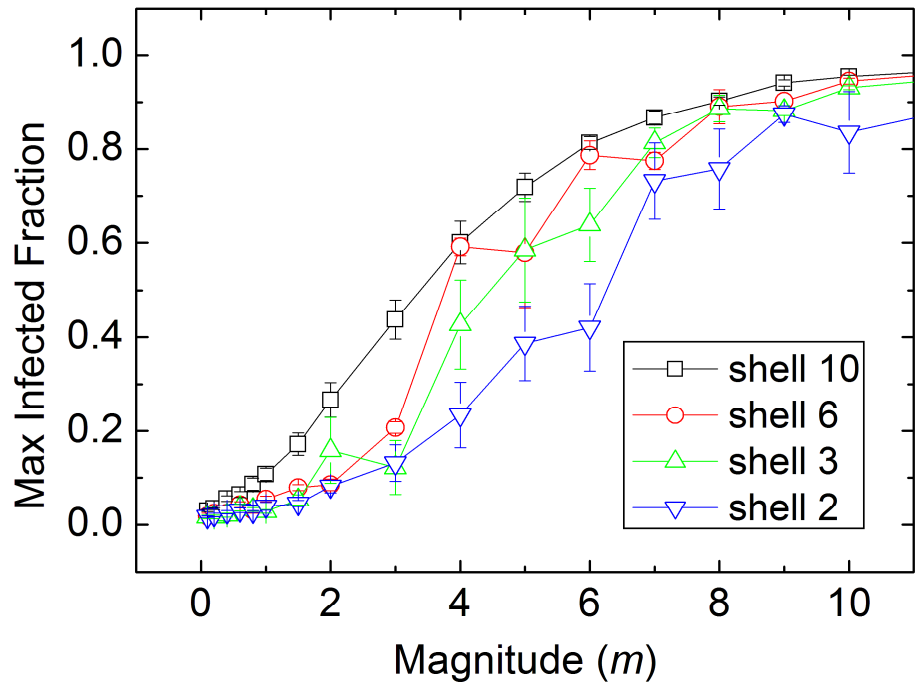

**Supplementary Figure S8**. The maximum fraction of the infected nodes by a crisis originating from different shells of the network versus its magnitude *m.* The results are averages over 50 realizations, and the error bars are showing the standard deviation.

# PART III

**Spreading of a crisis started in Greece.**

In this section we used our model to understand into what extent a crisis originating in Greece could spread to other countries. For both networks (CON and ITN) describing economic activity we obtained similar results. Our findings indicate that a large crisis has the potential to propagate to large percentage of the world countries. The worst case scenario, which is given by the maximum value of the infected fraction, shows a rapid global contagion for both networks, stressing that the current situation in Greece needs to be dealt fast and accurately.

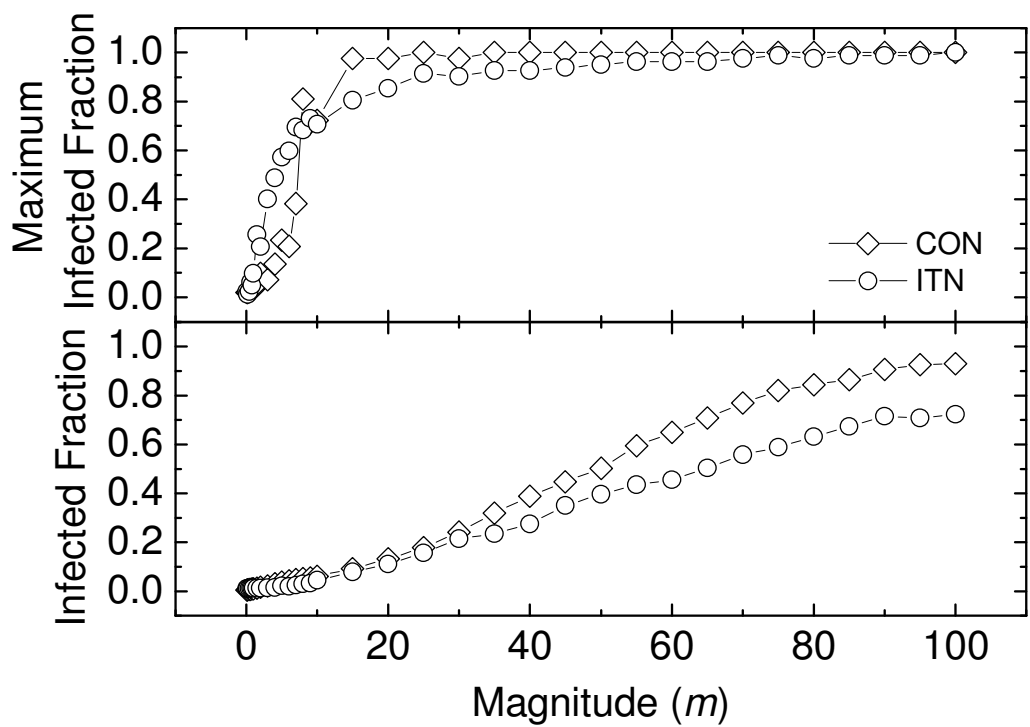

**Supplementary Figure S9**. The fraction of infected countries, and the maximum fraction of infected countries versus the magnitude *m* of a crisis originated in Greece and spread to the rest of the world through CON and ITN. The results are averages over 1000 realizations.

**Supplementary Table S1**. Mean probability of infection of different countries from crises originating in Greece for $m \in [1,100]$.

| Country code | CON | | ITN | | |
|---|---|---|---|---|---|
| | Mean infection probability | Standard deviation | Mean infection probability | Standard deviation | |
| AL | 0.97 | 0.17 | 0.2 | 0.40 | **Proximal countries** |
| MK | 0.98 | 0.14 | 0.22 | 0.41 | |
| BA | 0.86 | 0.34 | 0.34 | 0.47 | |
| TR | 0.59 | 0.49 | 0.38 | 0.48 | |
| IL | 0.41 | 0.49 | 0.31 | 0.46 | |
| BG* | 0.90 | 0.30 | 0.44 | 0.49 | |
| RO* | 0.82 | 0.38 | 0.38 | 0.48 | |
| PT | 0.45 | 0.50 | 0.38 | 0.48 | [*BG and RO are EU members as well] **EU members** |
| AU | 0.41 | 0.49 | 0.38 | 0.48 | |
| BE | 0.40 | 0.49 | 0.38 | 0.48 | |
| DE | 0.40 | 0.49 | 0.38 | 0.48 | |
| DK | 0.40 | 0.49 | 0.37 | 0.48 | |
| ES | 0.40 | 0.49 | 0.38 | 0.48 | |
| FI | 0.40 | 0.49 | 0.38 | 0.48 | |
| GB | 0.40 | 0.49 | 0.38 | 0.48 | |
| IT | 0.40 | 0.49 | 0.38 | 0.48 | |
| LU | 0.40 | 0.49 | 0.38 | 0.48 | |
| NL | 0.40 | 0.49 | 0.38 | 0.48 | |
| IE | 0.44 | 0.49 | 0.38 | 0.48 | |
| US | 0.40 | 0.49 | 0.38 | 0.48 | **Other large economies** |
| JP | 0.40 | 0.49 | 0.38 | 0.48 | |
| CA | 0.43 | 0.49 | 0.38 | 0.48 | |
| CN | 0.52 | 0.50 | 0.38 | 0.48 | |
| RU | 0.62 | 0.48 | 0.38 | 0.48 | |

As it is presented here, CON is more sensitive in comparison to the ITN, and it predicts an easier propagation of a crisis to countries in the neighbour of Greece (South-Eastern Europe). These countries traditionally have strong economic ties with Greece.



\end{document}